\documentclass[a4paper,11pt]{article}
\pdfoutput=1 
\usepackage{jinstpub} 
\usepackage{lineno}
\usepackage{siunitx}
\usepackage{graphicx}
\usepackage{subcaption}

\title{\boldmath CTPX1: A Highly Integrated and High-Throughput Data-Driven Camera Based on Timepix4}

\author[a,b,c]{Q. Li}
\author[a,b,1]{H. Liu\note{Corresponding authors: Hongbin Liu and Zhijia Sun}}
\author[a,b]{X. Jiang}
\author[a,b]{J. Zhou}
\author[e,f]{Y. Zhou}
\author[b,d]{H. Guo}
\author[a,b,c]{D. Cai}
\author[a,b,c]{W. Gong}
\author[a,b,c]{Y. Yuan}
\author[b,d]{C. Zhang}
\author[a,b]{S. Wang}
\author[a,b]{Y. Zhao}
\author[a,b,1]{Z. Sun}

\affiliation[a]{Institute of High Energy Physics, Chinese Academy of Sciences, Beijing, 100049, China}
\affiliation[b]{Spallation Neutron Source Science Center, Dongguan, Guangdong, 523803, China}
\affiliation[c]{University of Chinese Academy of Sciences, Beijing 100049, China}
\affiliation[d]{University of Electronic Science and Technology of China, Chengdu, Sichuan 610054, China}
\affiliation[e]{Purple Mountain Observatory, Chinese Academy of Sciences, Nanjing 210023, China}
\affiliation[f]{University of Science and Technology of China, Hefei 230026, China}
\emailAdd{hbliu@ihep.ac.cn;sunzj@ihep.ac.cn}

\abstract{
The upgrade of the China Spallation Neutron Source (CSNS-II) will raise the proton beam power from \qty{100}{\kilo\watt} to \qty{500}{\kilo\watt}, leading to a substantial rise in neutron flux. Consequently, the existing Timepix3-based detector systems, typically limited to a count rate of \qty{80}{Mhits/s}, will encounter severe saturation challenges.
To address the demand of the Energy-Resolved Neutron Imaging instrument (ERNI) for next-generation higher count-rate electronics, this paper presents CTPX1, a high-performance data-driven camera system developed based on the Timepix4 ASIC.
In terms of hardware design, the system adopts a compact modular architecture, deeply integrating readout electronics, precision high-voltage bias unit, and a TEC closed-loop temperature control subsystem.
Regarding firmware development, to fully exploit the readout potential of the Timepix4 ASIC's 16 high-speed serial links, this paper proposes a two-stage parallel processing and merging architecture. This architecture achieves real-time data aggregation with a total bandwidth of up to \qty{81.92}{\giga\text{bps}}. 
Experimental results demonstrate the camera's good long-term stability. Over a continuous 12-hour operation period, temperature fluctuations were kept within $0.1~^\circ\text{C}$ while the high-voltage output noise remained below $1~\text{mV}$.
High-flux X-ray testing indicates that the system achieves a peak event readout rate of  \qty{1.17}{\giga\text{hits}\per\second}, approaching the theoretical limit of the configured link speed. 
In-beam neutron verification at CSNS confirms that the camera successfully resolves fine spatial structures, achieving an imaging performance consistent with the \qty{55}{\micro\meter} pixel pitch of the sensor. Furthermore, the clear observation of spectral features in the Time of flight (TOF) spectrum of a $\gamma$-Fe sample validates the system's good time resolution.
This camera effectively addresses the data readout saturation challenges associated with the CSNS-II upgrade, validates the feasibility of Timepix4 technology for neutron imaging, and provides a viable solution for next-generation high-performance neutron imaging instrument.}


\begin{document}
\maketitle

\flushbottom
\section{Introduction}
\label{sec:introduction}
With the advancement of pulsed spallation neutron source technology, Energy Resolved Neutron Imaging (ERNI) has emerged as a pivotal non-destructive testing (NDT) technique in materials science for characterizing microstructures and stress distributions\cite{su2016time,Su_Oikawa_Shinohara_Kai_Horino_Idohara_Misaka_Tomota_2021,lehmann2014energy}. 
Leveraging the Time-of-Flight (ToF) method for neutron energy determination, ERNI delivers advanced imaging capabilities that integrate transmission imaging with the measurement of internal lattice strain\cite{shinohara2020energy,CHEN2024169460}. 
However, successful implementation of this imaging modality relies on a detector system that offers both micron-level spatial resolution to resolve fine structures and microsecond-level time resolution to ensure accurate ToF spectrum reconstruction.
\par
Within the detector array of the ERNI instrument at CSNS~\cite{yang2021novel,yang2024development}, the core detector responsible for high-precision energy-resolved imaging is the Tpx3Cam~\cite{fisher2016timepixcam,nomerotski2019imaging}, which is based on the Timepix3 ASIC~\cite{poikela2014timepix3}. 
Leveraging the data-driven architecture of Timepix3, this detector simultaneously records the position, Time-of-Arrival (ToA), and Time-Over-Threshold (ToT) of incident neutrons, having successfully achieved high spatial and temporal resolution neutron imaging during CSNS Phase I operations.
However, despite its current stability, the detector faces significant challenges due to the impending CSNS-II upgrade, where proton beam power will increase to \qty{500}{\kilo\watt}\cite{fu2019operation,Fu_2018}. This corresponds to an approximate five-fold increase in neutron flux. 
Under these conditions, the existing readout bandwidth of \qty{80}{\mega\text{hits}\per\second} approaches its saturation limit, which will inevitably lead to data pile-up and a marked increase in dead time, causing the loss of critical experimental data.
\par

To address the event rate saturation, we designed a high-throughput camera system based on the Timepix4 readout ASIC. Developed by the Medipix4 Collaboration, this chip serves as the core detection component of our design. 
As the successor to Timepix3, Timepix4 features a pixel array expanded to \numproduct{448 x 512}, offering an effective area approximately four times larger than its predecessor\cite{llopart2022timepix4,Heijhoff_2022}. 
The chip integrates 16 high-speed serializers. When configured at a link rate of \qty{5.12}{\giga\text{bps}} per lane, the total bandwidth reaches \qty{81.92}{\giga\text{bps}}.
Coupled with a peak event rate of \qty{1.25}{\giga\text{hits}\per\second}, these figures represent an improvement in count rate capability of over an order of magnitude compared to Timepix3.
This enhanced throughput capability effectively overcomes the bandwidth saturation issues within the high-flux environment of CSNS-II, thereby ensuring data integrity for the ERNI experiments.
Furthermore, the support for Through-Silicon Via technology in Timepix4 also establishes a hardware foundation for the future construction of large-field-of-view and seamless detector arrays.

\par
In this paper, we present the design and implementation of CTPX1, an event-driven camera system developed utilizing the Timepix4 ASIC. The device achieves efficient readout capabilities within a compact, integrated form factor. In terms of hardware, the system incorporates high-performance readout electronics, a custom low-noise high-voltage bias unit, and a precision closed-loop temperature control system. In terms of firmware, two-stage parallel processing and merging architecture is employed to enable the real-time aggregation of multiple high-bandwidth data streams. This study details the system architecture and design strategy of the camera. Finally, its readout performance, long-term stability, and imaging quality are evaluated through laboratory X-ray tests and in-beam neutron experiments at CSNS.

\section{System design and integration}
\label{sec:CTPX1 System Design and Integration}
\subsection{System Architecture}
\label{sec:System architecture}
In contrast to early electronics prototypes designed solely for functional verification, the CTPX1 system developed in this study is positioned as a general-purpose Timepix4 camera instrument.
To address operational requirements for portability and standalone operation, a compact system-level design is implemented to integrate the readout electronics with essential support circuits.
Through standardized mechanical and optical interface designs, the camera facilitates direct in-situ replacement of the existing Timepix3 camera systems currently installed at ERNI instrument\cite{CHEN2024169460}. 
This backward compatibility minimizes the engineering complexity and deployment costs of instrument system upgrades without altering existing mechanical constraints.
\par
\begin{figure}[htbp]
\centering
\includegraphics[width=0.8\textwidth]{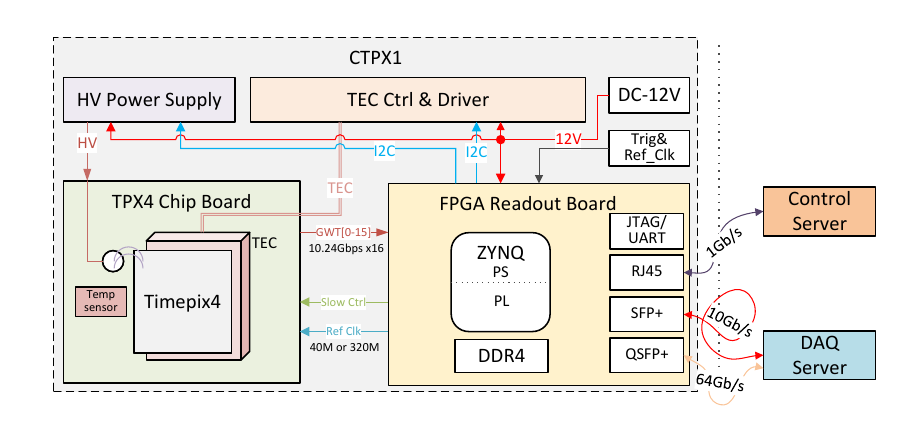}
\caption{Block diagram of the CTPX1 hardware architecture.\label{fig:CTPX1_Diagram}}
\end{figure}
\begin{figure}[htbp]
\centering
\includegraphics[width=0.7\textwidth]{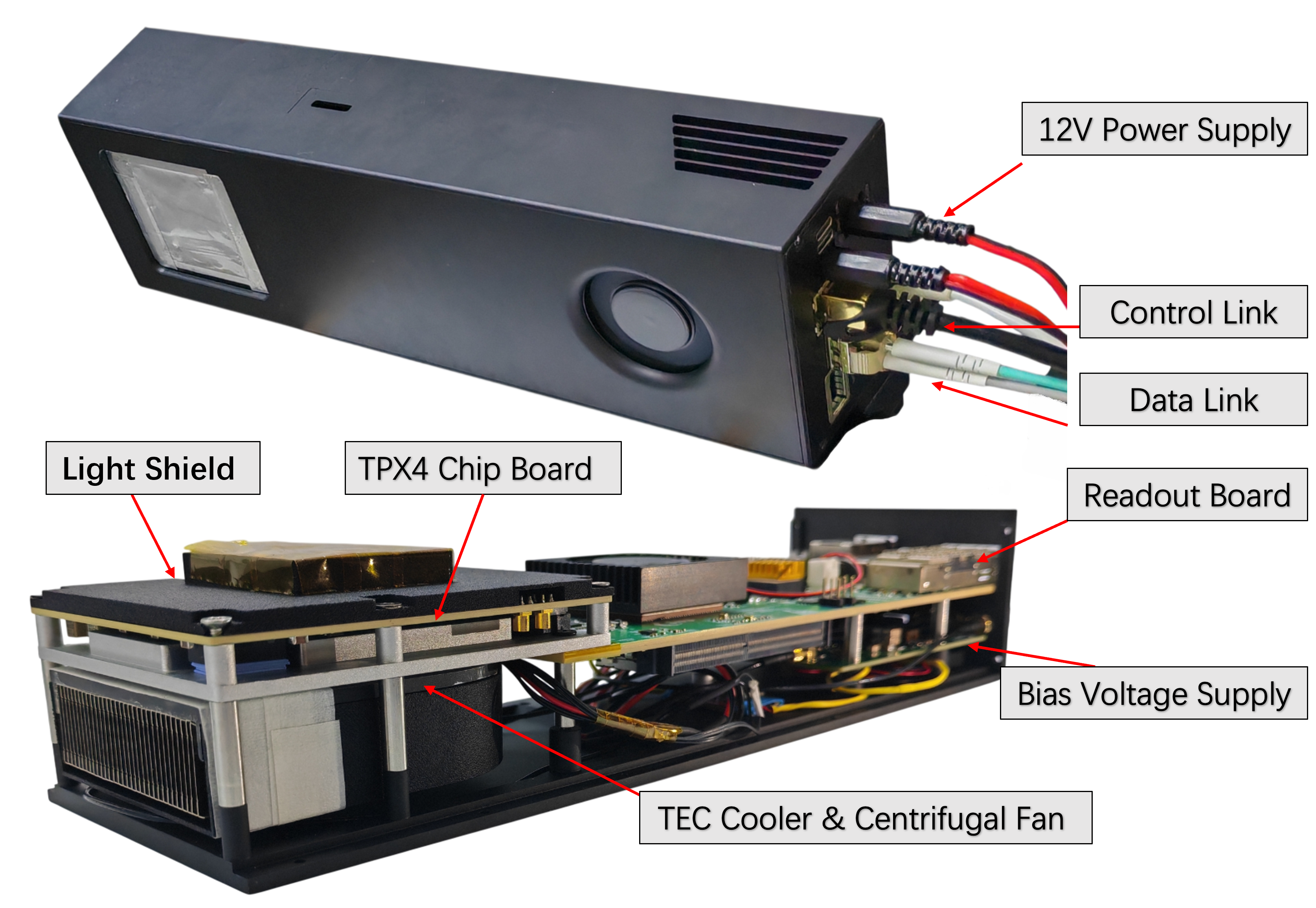}
\caption{Photograph of the assembled CTPX1 camera.}
\label{fig:CTPX1_Actual_Photos}
\end{figure}

To achieve these design objectives while balancing integration complexity with maintenance flexibility, CTPX1 employs a modular hardware architecture, as illustrated in Figure~\ref{fig:CTPX1_Diagram}.
The system adopts a distributed topology where the critical signal path is physically decoupled from the auxiliary support functions.
At the core of the architecture lies the FPGA-based readout carrier board, which serves as the primary controller responsible for managing the high-speed serial links from the ASIC and coordinating the operation of peripheral subsystems.
It interfaces with the chip board via a high-density connector to ensure signal integrity.
Surrounding this core, the high-voltage bias unit and the TEC-based thermal control module are implemented as independent sub-assemblies.
This modular approach allows for the coordinated operation of precise power delivery, thermal management, and high-speed data transmission, while facilitating individual module testing and replacement. A photograph of the CTPX1 Camera is presented in Figure~\ref{fig:CTPX1_Actual_Photos}.

\subsection{Hardware implementation}
\label{sec:Hardware System}
\subsubsection{Timepix4 readout electronics}
\label{sec:Timepix4 Readout Electronics}
The readout functionality of the CTPX1 camera is implemented using a custom electronics platform developed at CSNS(see \cite{li2025development} for detailed design specifications). 
This platform supplies low-noise power and synchronous clocks to the Timepix4 ASIC, configures the chip via a slow control interface, and manages the reception, aggregation, and backend transmission of 16 channels of high-speed serial data.
\par
In terms of physical implementation, the electronics comprises two primary modules: a Chip Board and an FPGA-based Readout Board. 
These components are interconnected via high-density FMC connectors, facilitating the stable transmission of power, clock signals, and the 16 high-speed data links.
The Chip Board accommodates a single Timepix4 assembly. This assembly comprises a \qty{300}{\micro\meter} thick $p^{+}$-in-$n$ silicon pixel sensor manufactured by Advafab Oy (Espoo, Finland)\cite{advafab_web}, which is flip-chip bonded to the underlying Timepix4 ASIC. 
Electrical connectivity between the assembly and the PCB is established through wire bonding.
\par
The FPGA Readout Board functions as the central processing unit of the electronics and utilizes the Xilinx Zynq UltraScale+ MPSoC\cite{mpsoc}. 
This device integrates a multi-core Processing System (PS) running the Linux operating system alongside a Programmable Logic (PL) unit.
The PL section features 24 GTH transceivers with maximum data rates of \qty{16.3}{\giga\text{bps}}.
These resources establish the hardware foundation for Timepix4 data readout and high-bandwidth backend transmission. To accommodate the varying event rate requirements of different experiments, the readout board incorporates high-capacity DDR4 memory as a data buffer. 
Additionally, the readout board offers multiple communication interfaces such as RJ45, SFP+, and QSFP+ to support data transmission rates ranging from \qty{1}{\giga\text{bps}} to \qty{64}{\giga\text{bps}}.
Leveraging the 16-channel readout capability established in previous work, the system successfully achieves stable operation at \qty{5.12}{\giga\text{bps}} per lane, yielding a maximum aggregated readout bandwidth of \qty{81.92}{\giga\text{bps}}.

\begin{figure}[htbp]
\centering
\includegraphics[width=1.0\textwidth]{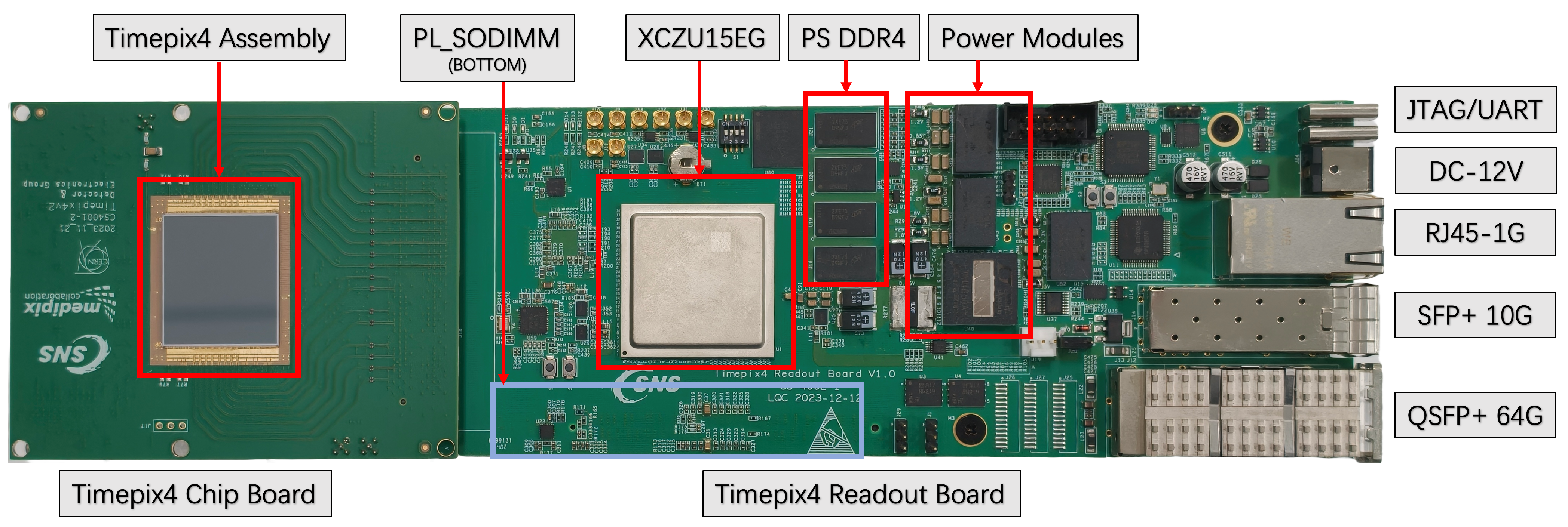}
\caption{Photograph of the single Timepix4 readout electronics.}
\label{fig:CTPX1_Readout_Electronics}
\end{figure}

\subsubsection{High voltage bias unit}
\label{sec:Sensor bias circuit}
\begin{figure}[htbp]
    \centering
    \begin{subfigure}[b]{0.48\textwidth}
        \centering
        \includegraphics[width=\textwidth]{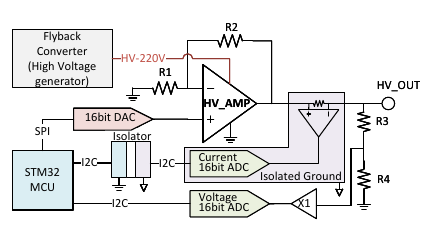}
        \caption{}
        \label{fig:CTPX1_HV_Diagram}
    \end{subfigure}
    \hfill 
    \begin{subfigure}[b]{0.48\textwidth}
        \centering
        \includegraphics[width=\linewidth]{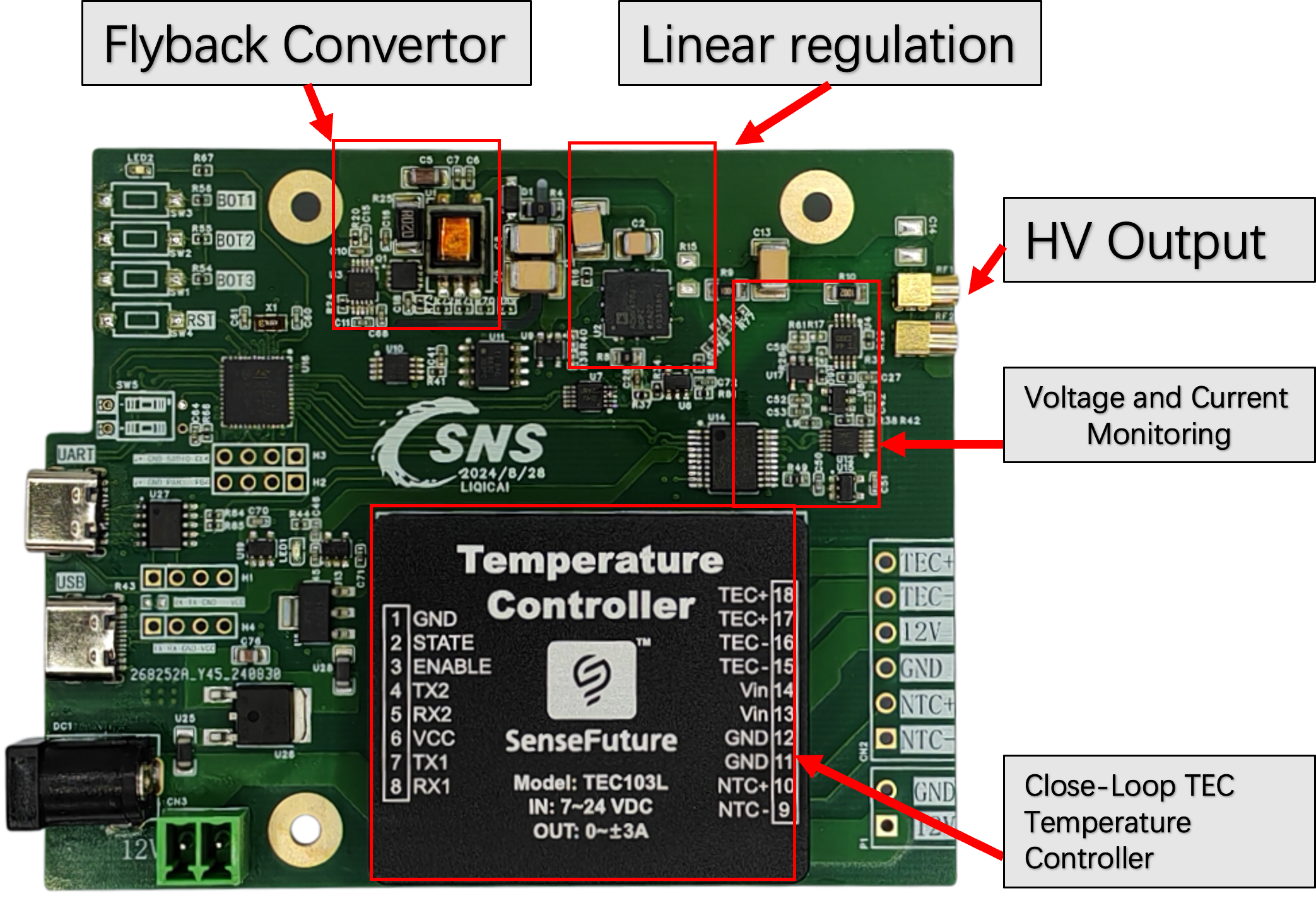} 
        \caption{}
        \label{fig:HV_module_picture}
    \end{subfigure}
    \caption{
    (a) Block diagram of the high-voltage bias unit.
    (b) Photograph of the high-voltage bias unit and temperature controller board.}
    \label{fig:hv_picture}
\end{figure}

To ensure the \qty{300}{\micro\meter} silicon pixel sensor operates at full depletion, which is essential for optimizing charge collection efficiency and time resolution, the CTPX1 system integrates a dedicated high-voltage bias module.
Providing the same programmable bias voltage and real-time monitoring capabilities as bulky external supply units, this module implementation is distinguished by its compact footprint.
Figure~\ref{fig:CTPX1_HV_Diagram} presents the schematic block diagram of the HV module, which employs a two-stage hybrid architecture.
The first stage functions as a pre-regulator utilizing a Flyback topology to boost the \qty{12}{\volt} system input to a DC voltage of approximately \qty{220}{\volt}. The second stage operates as a linear regulator for final voltage adjustment and noise filtering.
In this stage, a microcontroller controls a \qty[quantity-product = -]{16}{\bit} DAC to generate a \qtyrange{0}{5}{\volt} reference voltage.
This reference signal drives a high-voltage operational amplifier configured in a non-inverting topology to perform linear regulation\cite{ADHV4702_datasheet}.
Such a configuration facilitates high-voltage control via low-voltage signals and suppresses ripple noise introduced by the switching stage.
To facilitate status monitoring, the circuit employs a resistive divider network for voltage readback and a high-side sensing topology to measure leakage currents. 
These readback signals are transmitted to the microcontroller unit for processing. Through embedded digital control algorithms, the system implements programmable voltage ramping alongside over-voltage and over-current protection mechanisms to ensure the operational safety of the sensor module. 

\subsubsection{Temperature control subsystem}
\label{sec:TEC Closed-Loop}
The typical power consumption of the Timepix4 readout ASIC ranges from \qty{2}{\watt} to \qty{5}{\watt}\cite{llopart2022timepix4}. The resulting heat dissipation induces an increase in sensor leakage current, which restricts the maximum applicable bias voltage. To address the integration and portability requirements of the CTPX1 system, a thermal management strategy combining a Thermoelectric Cooler (TEC) and forced air cooling is implemented. This approach obviates the need for the liquid circulation infrastructure associated with water cooling systems. Structurally, the hot side of the TEC module connects to a turbine heatsink assembly. The cold side interfaces with the rear of the Chip Board using a thermal interface material to establish a low-resistance thermal conduction path. To ensure precise thermal regulation, a commercial PID closed-loop temperature control module~\cite{SenseFuture_TEC103L} is integrated into the camera. This module dynamically adjusts the TEC drive current to suppress temperature fluctuations, maintaining the sensor at a stable operating setpoint.

\subsection{Firmware and software design}
\label{sec:Firmware and Software Design}
\subsubsection{Firmware design}
\label{sec:Firmware Design}
\begin{figure}[htbp]
\centering
\includegraphics[width=1.0\textwidth]{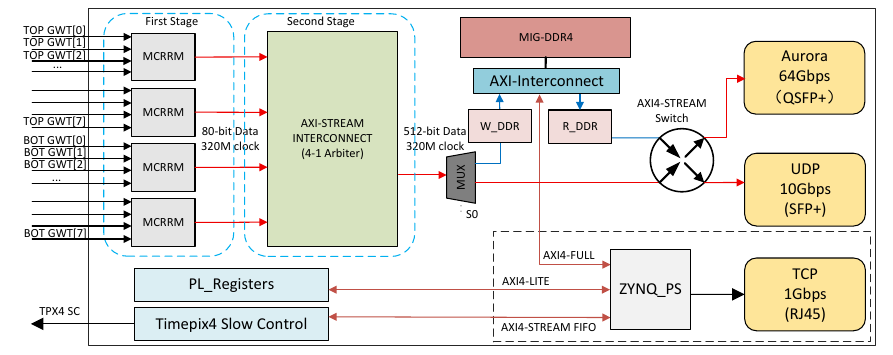}
\caption{Block diagram of the CTPX1 firmware.\label{fig:Firmware_Diagram}}
\end{figure}

The Timepix4 readout ASIC supports parallel readout via 16 GWT high-speed serial links.
A primary challenge in firmware design involves the efficient handling of these concurrent data streams within the FPGA logic.
Specifically, this requires descrambling, parallel processing, and non-blocking aggregation of all 16 signal channels.
To address these requirements, the CTPX1 readout system employs the firmware architecture illustrated in Figure~\ref{fig:Firmware_Diagram}.
\par
Unlike early firmware iterations limited to functional verification, the current design expands the number of active readout links from 2 to 16.
Simultaneously, the data rate per link increases from \qty{2.56}{\giga\text{bps}} to \qty{5.12}{\giga\text{bps}}.
In terms of flexibility, the firmware design allows for the independent configuration and masking of each link to accommodate varying bandwidth requirements.
Furthermore, the firmware implements a “Streaming Mode” for continuous real-time data transmission, alongside a “Buffered Mode” designed to absorb high-speed data bursts exceeding uplink capacity.
\par
As illustrated in Figure~\ref{fig:Firmware_Diagram}, the firmware architecture is implemented on the Xilinx Zynq UltraScale+ platform. 
The PS hosts an embedded operating system responsible for managing PL register configuration and the slow control of the Timepix4 ASIC via the AXI bus. Additionally, it provides a TCP interface for the remote management and status monitoring of the readout system. 
Within the PL, the data path employs a two-stage parallel processing and merging architecture to ensure efficient aggregation.
In the first stage, the Gigabit Transceiver H-series (GTH) modules in the FPGA concurrently acquire data streams from the 16 high-speed Gigabit Wire Transmitter (GWT) serial links distributed across the top and bottom halves of the Timepix4.
These data streams are processed by four Multi-Channel Round-Robin Merger (MCRRM) modules, which perform initial descrambling, timestamp extension, and 4:1 intra-group aggregation. 
This stage produces four formatted data streams, each with an 80-bit width operating at \qty{320}{\MHz}.
Subsequently, these streams are routed into a global AXI-Stream Interconnect for the second stage of global merging and bit-width conversion.
This process ultimately yields a \qty[quantity-product = -]{512}{\bit} wide AXI-Stream bus.
\par
\begin{figure}[htbp]
\centering
\includegraphics[width=1.0\textwidth]{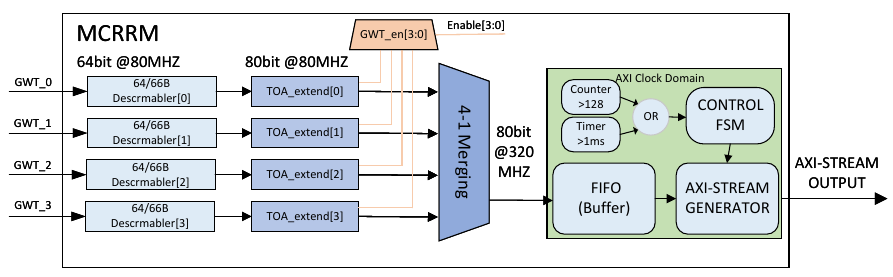}
\caption{Schematic of the multi-channel round-robin merger logic.\label{fig:Firmware_MCRRM}}
\end{figure}

Figure~\ref{fig:Firmware_MCRRM} details the internal architecture of the MCRRM. 
Upon entry, incoming data streams undergo descrambling to restore the 64-bit raw format. Subsequently, a timestamp extension logic expands the 16-bit coarse TOA timestamp to 32 bits, yielding a final event width of 80 bits. 
This extension prevents counter wraparound during long-duration acquisitions.
To implement the 4:1 aggregation, the module employs a clock-frequency boosting strategy. 
By increasing the processing clock from the link-side frequency of \qty{80}{\MHz} to \qty{320}{\MHz}, the logic achieves non-blocking, zero-loss aggregation of four input channels into a single pipeline. Given the inherent sparsity of Timepix4 event data, the merger integrates a first-level buffer governed by a flow control state machine. 
This logic utilizes a dual-trigger mechanism to regulate packet transmission. First, a counts-driven trigger initiates a burst transfer when the valid event count reaches a preset threshold (128) to maximize bus utilization. 
Second, a latency-driven trigger forcibly asserts the TLAST signal to flush buffered data if the wait time exceeds a defined timeout (\qty{1}{\ms}). 
Collectively, this traffic shaping design eliminates invalid clock cycles, reduces bus idle overhead, and enhances the arbitration efficiency of the subsequent global merger.
\par
To facilitate flexible data routing, the firmware integrates AXI4-Stream Switch and MUX logic designed to support two distinct operating modes. In "Streaming Mode", data streams bypass the memory buffer and are directed immediately to the output interfaces. This configuration enables real-time transmission via either the \qty{64}{\giga\text{bps}} Aurora protocol over QSFP+ or the \qty{10}{\giga\text{bps}} UDP protocol over SFP+.
Conversely, in "Buffered Mode", data streams are routed through memory controller modules into the onboard \qty{8}{GB} DDR4 memory. This design effectively decouples acquisition from transmission, allowing the system to buffer short-duration high-speed bursts at the full bandwidth of \qty{81.92}{\giga\text{bps}} (up to \qty{0.5}{\s}). Once the burst is captured, the data is then read out from the memory for downstream transmission.

\subsubsection{Software framework}
\label{sec:Software Framework}
In the current experimental setup, a single server hosts both system control and data acquisition services.
Accordingly, the software suite comprises two distinct subsystems: control and data acquisition. 
The control software framework follows a two-layer architecture consisting of the embedded system software on the Zynq Processing System and the host-side application. 
To reduce architectural complexity, the embedded software functions as a transparent bridge between the remote host and the hardware abstraction layer. 
Developed in C++ within a customized operating environment, this module encapsulates fundamental hardware operations, including Timepix4 register configuration, high-voltage bias adjustment, and sensor status monitoring. 
On the host side, system operation relies on a Python-based Command Line Interface, with a graphical user interface currently under development.
For the data acquisition subsystem, the software supports continuous data reception and storage via either standard \qty{10}{GbE} network interfaces or dedicated PCIe cards utilizing the Aurora protocol. 
As the current iteration prioritizes maximum raw data throughput to disk, real-time parsing and online post-processing capabilities are not implemented.

\begin{figure}[htbp]
\centering
\includegraphics[width=1.0\textwidth]{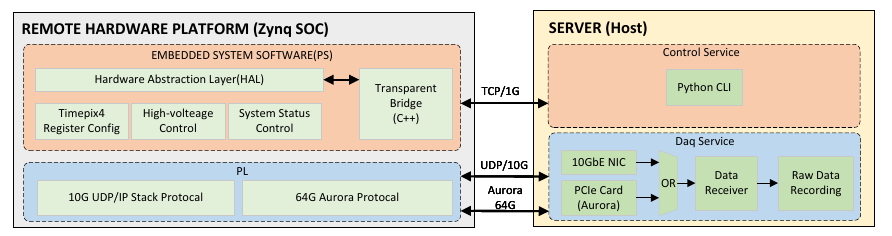}
\caption{The software framework of CTPX1 camera}
\label{fig:Software_Framework}
\end{figure}

\section{Camera system evaluation}
\label{fif:tests}

Following the hardware fabrication and firmware integration of the CTPX1 camera system, a series of system-level verification tests were conducted to evaluate its performance.
The evaluation process began with the characterization of critical support subsystems, specifically focusing on the regulation stability of the TEC thermal control module and the linearity and noise performance of the high-voltage bias unit.
Subsequently, a throughput stress test was performed using a high-flux X-ray source to determine the bandwidth limits of the readout electronics. 
Finally, the system was deployed at a neutron beamline facility to validate its imaging capabilities under experimental conditions.

\subsection{Camera temperature stability test}
\label{sec:temperature_tests}

The power consumption of the Timepix4 ASIC varies dynamically across different operating modes.
For precision pixel detectors, maintaining long-term temperature stability is often more critical than achieving the lowest possible absolute temperature, as thermal fluctuations directly affect detector gain consistency and noise.
To evaluate the capability of the CTPX1 thermal management system in handling dynamic load changes, the integrated TEC closed-loop control system was subjected to a dynamic response test.
In this experiment, temperature data was acquired in real-time at a sampling rate of \qty{1}{\hertz} using a sensor mounted on the PCB backside directly underneath the Timepix4 ASIC, with the target temperature set to \qty{25}{\degreeCelsius}. 
\par
\begin{figure}[htbp]
    \centering
    \begin{subfigure}[b]{0.48\textwidth}
        \centering
        \includegraphics[width=\textwidth]{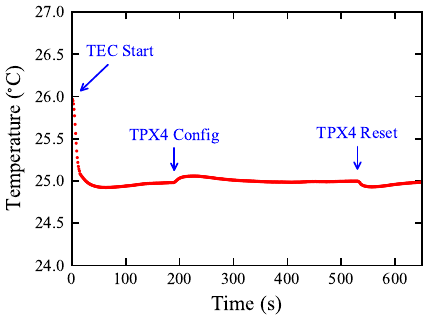}
        \caption{}
        \label{fig:temp_Dynamic_Response}
    \end{subfigure}
    \hfill 
    \begin{subfigure}[b]{0.48\textwidth}
        \centering
        \includegraphics[width=\linewidth]{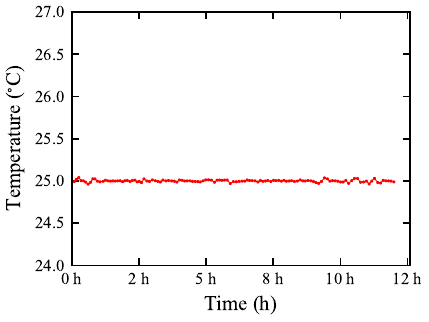} 
        \caption{}
        \label{fig:temp_stability}
    \end{subfigure}
    \caption{Temperature performance of the CTPX1 camera system. (a) Dynamic temperature response during the initialization sequence. The annotated events correspond to TEC activation, chip configuration, and reset phases. (b) Long-term temperature stability monitoring over a 12-hour period under steady-state operating conditions.}
    \label{fig:temp_picture}
\end{figure}

Figure~\ref{fig:temp_Dynamic_Response} illustrates the temperature response curve throughout the startup sequence. 
The annotated events correspond to TEC activation, register configuration, and chip reset.
The results indicate that following the event labeled "TEC Start," the controller responded rapidly, cooling the chip from ambient temperature to the target setpoint.
During the subsequent "TPX4 Config" and "TPX4 Reset" phases, internal logic state transitions typically trigger transient power variations.
However, the monitoring data shows that even during these operations, the fluctuation in temperature was effectively contained within \qty{0.1}{\degreeCelsius}. No significant thermal drift was observed, demonstrating the good regulation capability of the system.
To further assess the system reliability during long-duration experiments, a 12-hour continuous temperature monitoring test was conducted under normal operating conditions.
Figure~\ref{fig:temp_stability} presents the temporal temperature profile recorded during this period.  
The data reveals that throughout the run, the peak-to-peak temperature fluctuation was also consistently maintained within \qty{0.1}{\degreeCelsius}. This result verifies that the thermal control system exhibits good stability.

\subsection{Characterization of the CTPX1 high voltage unit}
\label{sec:Bias_hv_test}
\begin{figure}[htbp]
\centering
\includegraphics[width=0.8\textwidth]{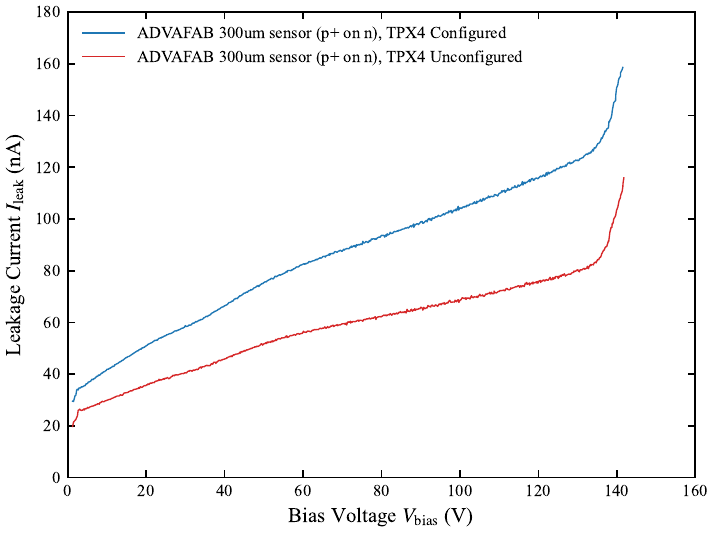}
\caption{I-V curves of the Timepix4 hybrid pixel detector with an ADVAFAB \qty{300}{\micro\meter} p+-on-n sensor. The blue line indicates the leakage current in the configured state, while the red line indicates the unconfigured state.}
\label{fig:Sensor_iv_curve}
\end{figure}

A stable and low-noise high-voltage bias is a prerequisite for the reliable operation of hybrid pixel detectors.
\begin{figure}[htbp]
    \centering
    \begin{subfigure}[b]{0.48\textwidth}
        \centering
        \includegraphics[width=\linewidth]{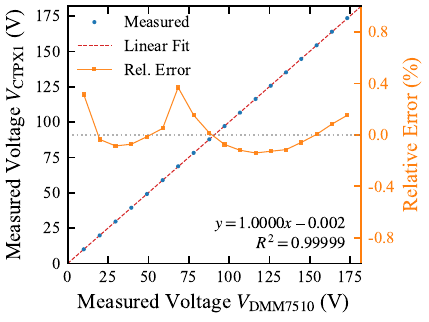}
        \caption{}
        \label{fig:HV_Bias_analysis}
    \end{subfigure}
    \hfill
    \begin{subfigure}[b]{0.48\textwidth}
        \centering
        \includegraphics[width=\linewidth]{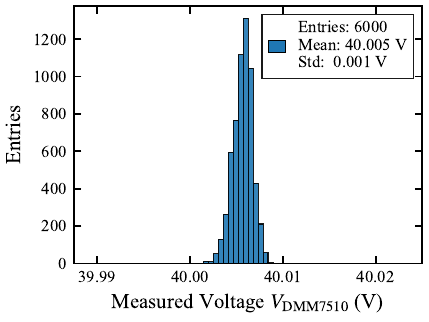}
        \caption{}
        \label{fig:HV_Bias_Noise}
    \end{subfigure}
    \caption{Performance characterization of the CTPX1 HV module. 
    (a) The measured curve showing the linearity between the Keithley DMM7510 instrument and the CTPX1 HV module, with relative error plotted on the right axis. 
    (b) The statistical distribution of the voltage measured at a fixed \qty{40}{\volt} output.}
    \label{fig:HV_picture}
\end{figure}
Prior to integrating the HV module into the camera, its linearity was calibrated and its noise performance was evaluated using a 7.5-digit digital multimeter (DMM7510, Keithley)\cite{Keithley_DMM7510}. 
Figure~\ref{fig:HV_Bias_analysis} presents the comparison between the calibrated internal ADC readback values and the reference measurements.
The calibrated voltage readback demonstrates high linearity across the measured range of \qtyrange{0}{175}{\volt} ($R^2 > \num{0.999}$), with the relative error magnitude consistently maintained within \qty{0.5}{\percent}.
Furthermore, at an operating point of \qty{40}{V}, the measured RMS noise voltage was approximately \qty{1}{\milli\volt}.
These metrics confirm that the HV module possesses the precision and low-noise characteristics necessary to satisfy the bias requirements of the Timepix4 sensor.
Utilizing the calibrated HV module, an IV characterization was performed on the sensor assembly to determine its optimal operating bias range.
The test was conducted at a regulated temperature of \qty{25}{\degreeCelsius}, sweeping the bias voltage from \qtyrange{0}{150}{\volt} in \qty{1}{\volt} increments.
Figure~\ref{fig:Sensor_iv_curve} illustrates the leakage current response curves for the Timepix4 in both unconfigured and configured states.
Analysis of the curves reveals that in the initial voltage range ($V < \qty{60}{\volt}$), the leakage current exhibits a characteristic rise associated with depletion zone expansion, yet remains well-contained below \qty{100}{\nano\ampere}. Within the \qtyrange{60}{135}{\volt} interval, the current transitions into a stable linear behavior, indicative of the sensor operating in its fully depleted regime. Beyond \qty{135}{\volt}, the leakage current rises sharply, signaling the onset of sensor breakdown. 
It is observed that the breakdown voltage threshold is preserved, although the overall leakage current is higher when the Timepix4 matrix is configured. This observation provides a reference for establishing bias protection thresholds in experimental setups.

\subsection{Readout bandwidth and high X-Ray flux response}
\label{sec:Readout_Bandwidth_Limits}

\begin{figure}[htbp]
\centering
\includegraphics[width=0.48\textwidth]{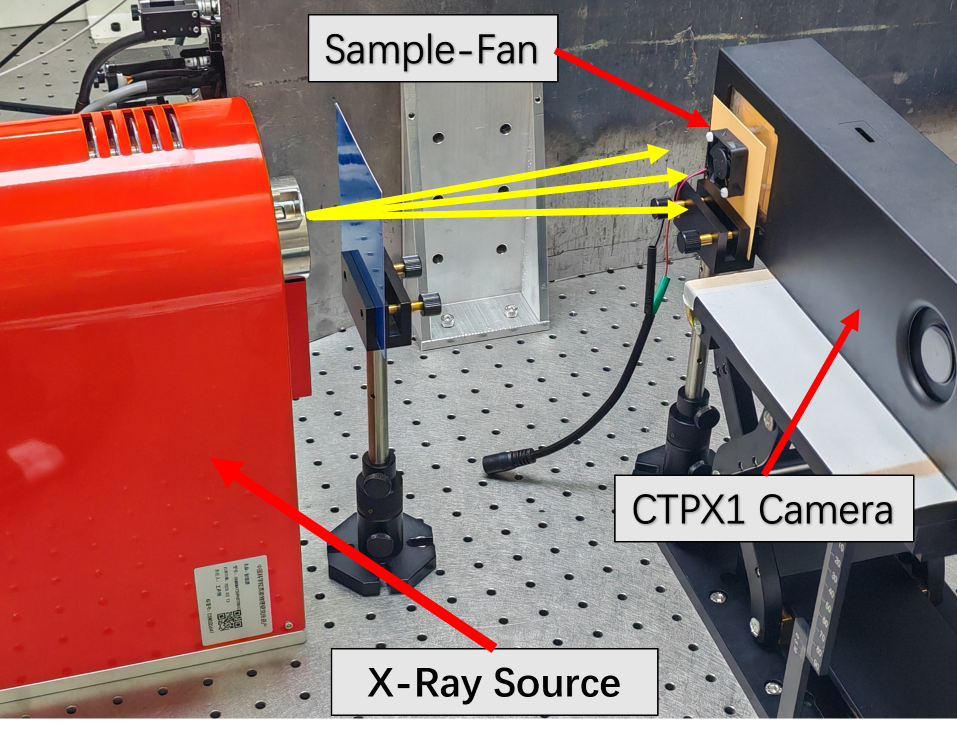}
\caption{X-ray Testing Environment.The detector was positioned $30~\text{cm}$ from the source to ensure the sensor was fully illuminated by the X-ray cone beam. To avoid leakage current induced by ambient visible light, the sensor area was shielded with a $100~\mu\text{m}$ thick aluminum foil. Throughout the experiments, the sensor bias voltage was maintained at $40~\text{V}$, and the X-ray tube voltage was fixed at $70~\text{kV}$.}
\label{fig:X-Ray test Environment}
\end{figure}

Dedicated readout electronics and firmware were developed to support the parallel operation of all sixteen \qty{5.12}{\giga\text{bps}} GWT links, designed to fully leverage the high bandwidth potential of the Timepix4 ASIC.
To experimentally validate the throughput limits of the CTPX1 camera, a high-flux test platform was established using a micro-focus X-ray source.
The experimental setup is illustrated in Figure~\ref{fig:X-Ray test Environment}.
A micro-focus X-ray source (L12161-07, Hamamatsu Photonics) \cite{hamamatsu} served as the high-flux irradiation device.
With a maximum tube voltage of \qty{150}{\kilo\volt} and a peak tube current of \qty{500}{\micro\ampere}, this device allows for the generation of photon fluxes across a wide dynamic range.
To evaluate the imaging performance under high count-rate conditions, a continuously rotating fan was introduced as a dynamic test target.
This object obstructed approximately \qty{50}{\percent} of the field of view, creating a high-contrast dynamic test scenario.
\par
Prior to the measurements, threshold equalization was performed across the full pixel array in Photon Counting 24-bit mode using the noise-edge method.
A global threshold was set to \qty{1}{\kilo\electronvolt}, and approximately 150 noisy pixels were masked.
For the experimental data acquisition, the Timepix4 ASIC was configured in data-driven mode and the acquisition duration was precisely controlled at \qty{0.5}{\second} via the Timepix4 Shutter signal. 
During the measurements, the generated data stream was buffered in the onboard DDR4 memory prior to readout.
This buffering strategy effectively avoids the bandwidth limitations of the backend QSFP+ interface (maximum $64~\text{Gbps}$) during peak flux, ensuring that the throughput is constrained only by the physical bandwidth of the 16 GWT links.
\par
In this study, the sensor current was utilized as the reference metric for incident photon flux intensity. 
This metric was chosen because it linearly reflects the incident flux and remains unaffected by the bandwidth limitations or dead time of the readout electronics.
\begin{figure}[htbp]
\centering
\includegraphics[width=0.6\textwidth]{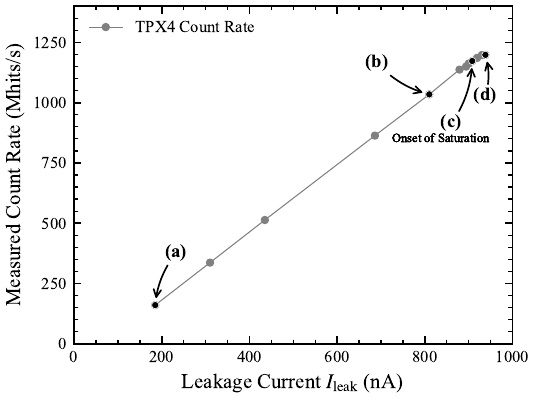}
\caption{Measured Timepix4 count rate versus sensor current under high-flux X-ray irradiation. The sensor leakage current (X-axis) serves as a proxy for the incident X-ray flux. The annotated points (a),(b),(c),(d) correspond to the count rates of the respective 2D imaging results shown in Figure~\ref{fig:Fan_img}.}
\label{fig:count_rate}
\end{figure}
\begin{figure}[htbp]
\centering
\includegraphics[width=0.7\textwidth]{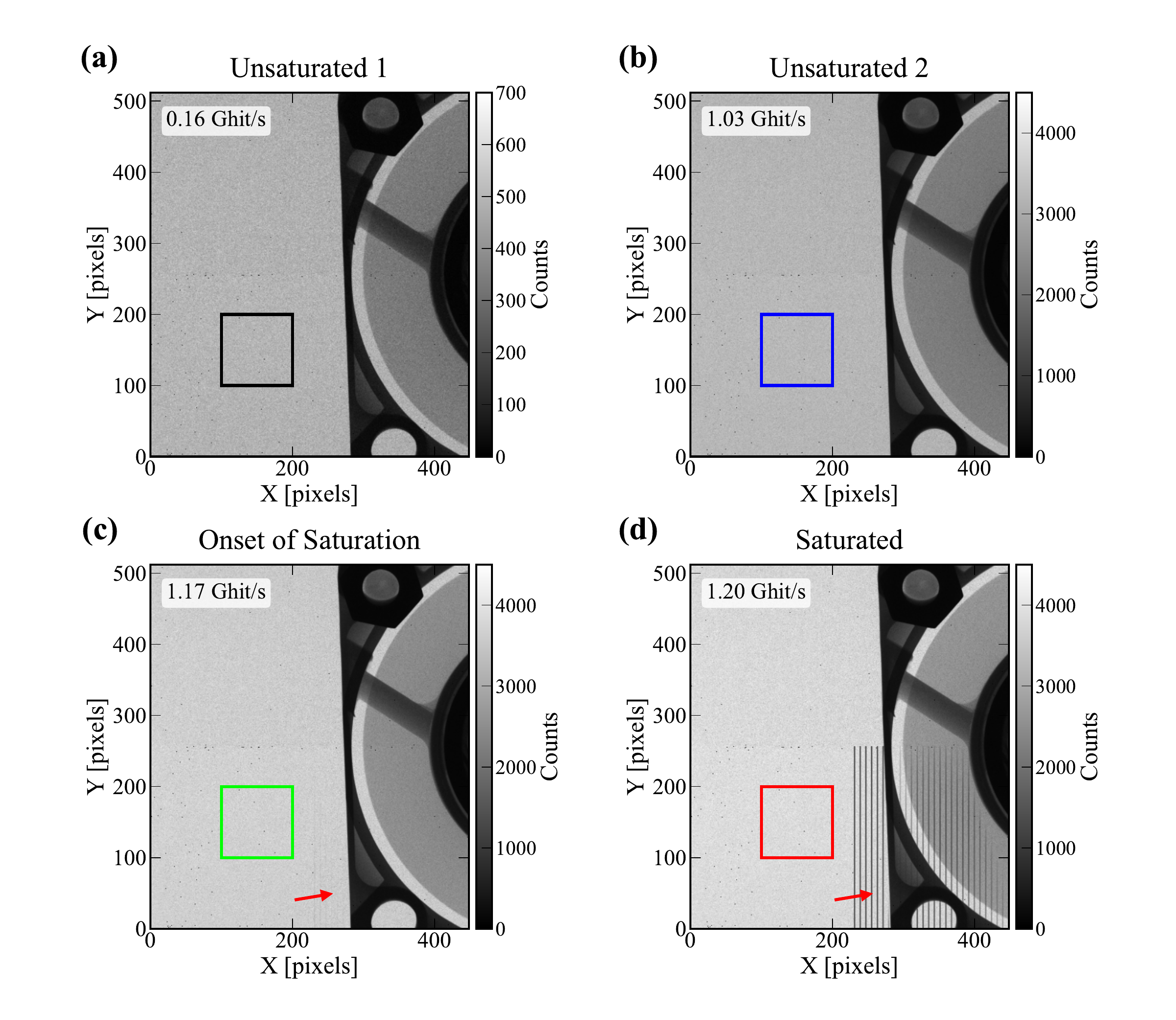}
\caption{2D X-ray imaging results acquired at global count rates of (a) \qty{0.16}{GHits/s}, (b) \qty{1.03}{GHits/s}, (c) \qty{1.17}{GHits/s}, and (d) \qty{1.20}{GHits/s}. Vertical striations indicating data loss begin to appear in the bottom-right region of (c) and become significantly more pronounced in (d). The rectangular boxes indicate the Regions of Interest (ROIs) used to extract the energy spectra presented in Figure~\ref{fig:tot_spec}.}
\label{fig:Fan_img}
\end{figure}
Figure~\ref{fig:count_rate} illustrates the relationship between the recorded count rate of the CTPX1 camera and the sensor current across varying flux levels. 
Figure~\ref{fig:Fan_img} presents the imaging results acquired at four distinct count rates.
The results demonstrate that in the low-flux regime, the output count rate scales linearly with the sensor current (i.e., incident flux).  
However, saturation effects emerge as the input event rate approaches the bandwidth limit of the 16 GWT links, specifically at an event rate of approximately \qty{1.17}{Ghits/s}.
At this point, vertical striation artifacts appear in the images, indicating data loss. As the tube current is further increased, these data loss artifacts become more pronounced, and the maximum recorded count rate plateaus at a stable value of \qty{1.20}{Ghits/s}, independent of further increases in input flux.
\par
\begin{figure}[htbp]
    \centering
    \begin{subfigure}[b]{0.48\textwidth}
        \centering
        \includegraphics[width=\linewidth]{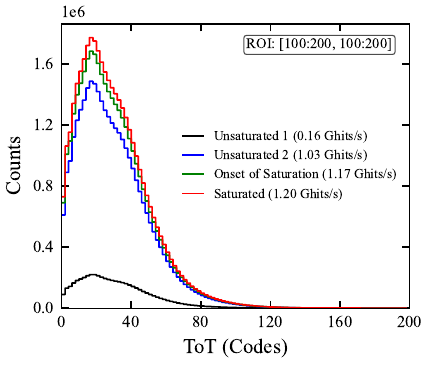} 
        \caption{}
        \label{fig:TOT_spectrum}
    \end{subfigure}
    \begin{subfigure}[b]{0.48\textwidth}
        \centering
        \includegraphics[width=\linewidth]{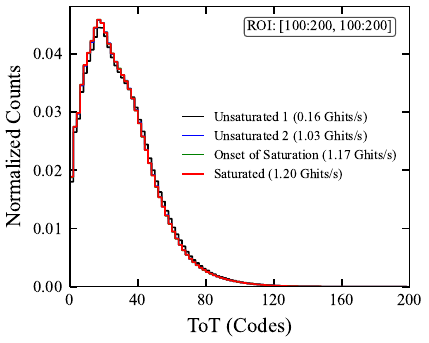} 
        \caption{}
        \label{fig:TOT_spectrum_norm}
    \end{subfigure}
    \caption{ToT energy spectra within the Region of Interest (ROI) at different count rates. (a) Unnormalized ToT spectra measured at four distinct count rates. (b) Normalized ToT spectra for the same four count rates.}
    \label{fig:tot_spec}
\end{figure}

To evaluate data integrity under high-flux conditions, we analyzed the ToT spectra extracted from a region of interest (ROI: Rows 100–200, Cols 100–200) across the four flux scenarios presented in Figure~\ref{fig:Fan_img}.
As illustrated in Figure~\ref{fig:tot_spec}, the normalized ToT spectral profile remains virtually identical across all flux rates.
This consistency confirms that the analog front-end maintains linearity under high load, suggesting that the saturation is strictly digital in nature, caused by the bandwidth limits of the GWT links and subsequent readout electronics.
These results confirm that the CTPX1 camera achieves a maximum lossless count rate of \qty{1.17}{Ghits/s}, utilizing approximately \qty{93}{\percent} of the theoretical bandwidth of the sixteen GWT links( \qty{1.25}{Ghits/s}).

\subsection{Neutron imaging verification}
\label{sec:Neutron Imaging}
\begin{figure}[htbp]
    \centering
    \begin{subfigure}[b]{0.48\textwidth}
        \centering
        \includegraphics[width=\linewidth]{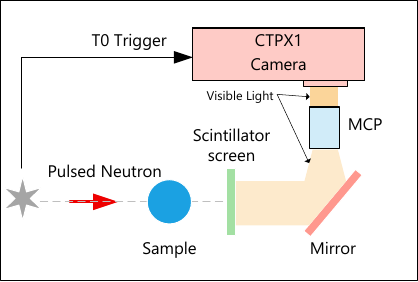} 
        \caption{}
        \label{fig:Neutron_Imaging_Detector}
    \end{subfigure}
    \begin{subfigure}[b]{0.45\textwidth}
        \centering
        \includegraphics[width=\linewidth]{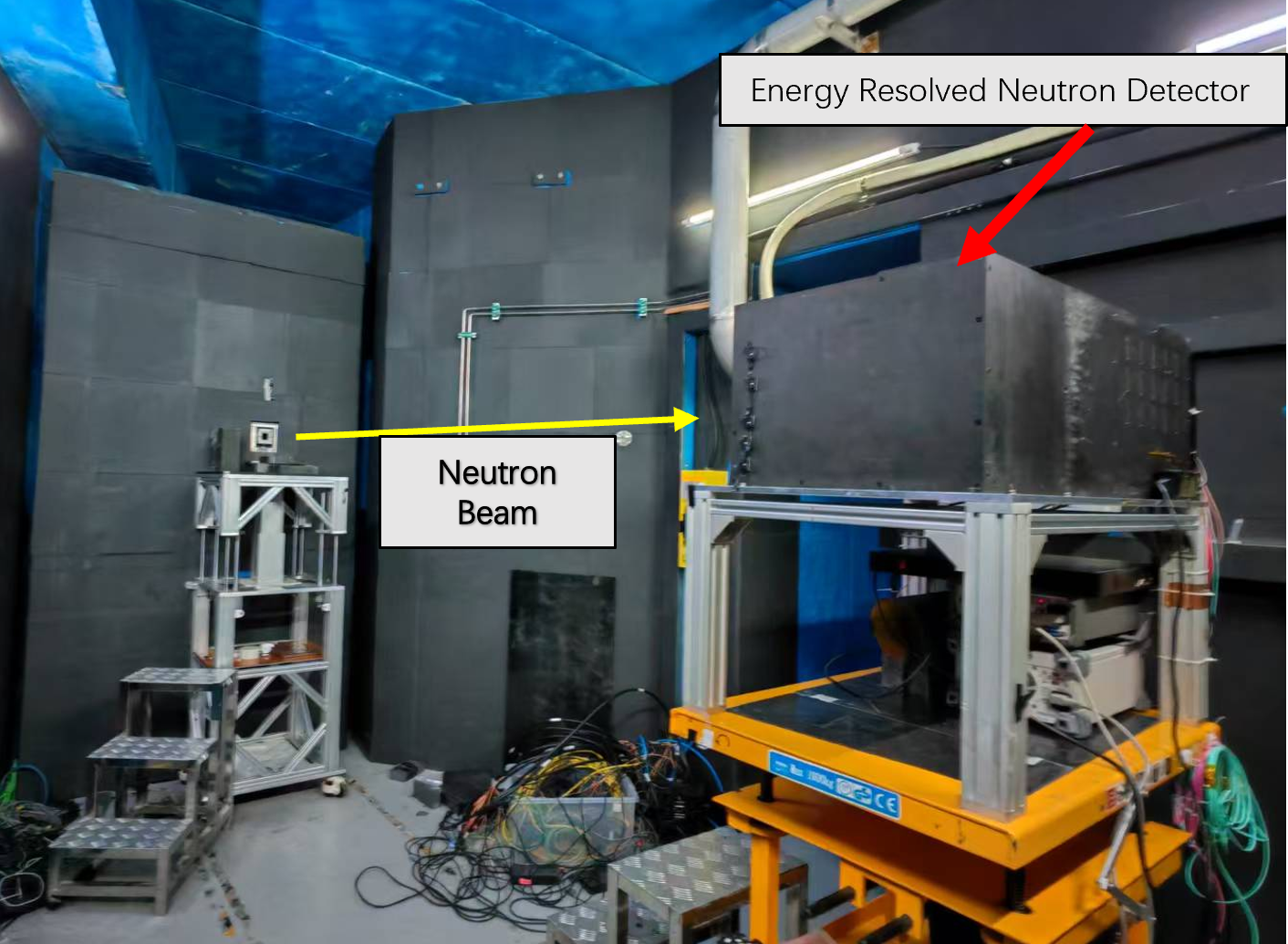} 
        \caption{}
        \label{fig:Neutron_Test_Environment}
    \end{subfigure}
    \caption{
    (a) Schematic diagram of the energy-resolved neutron imaging principle.
    (b) Photograph of the test environment installed at the BL20 beamline of CSNS.
    }
    \label{fig:neutron_test}
\end{figure}

Performance verification experiments were conducted at the BL20 beamline of CSNS to evaluate the CTPX1 camera in neutron imaging applications.
The primary objective was to validate the system's capabilities in two-dimensional spatial resolution and neutron ToF spectrometry.
A schematic illustrating the experimental principle is presented in Figure~\ref{fig:Neutron_Imaging_Detector}. For a more detailed description of the detector's working principle and specifications, please refer to Ref.~\cite{yang2021novel}.\par
\begin{figure}[htbp]
    \centering
    \includegraphics[width=0.4\textwidth]{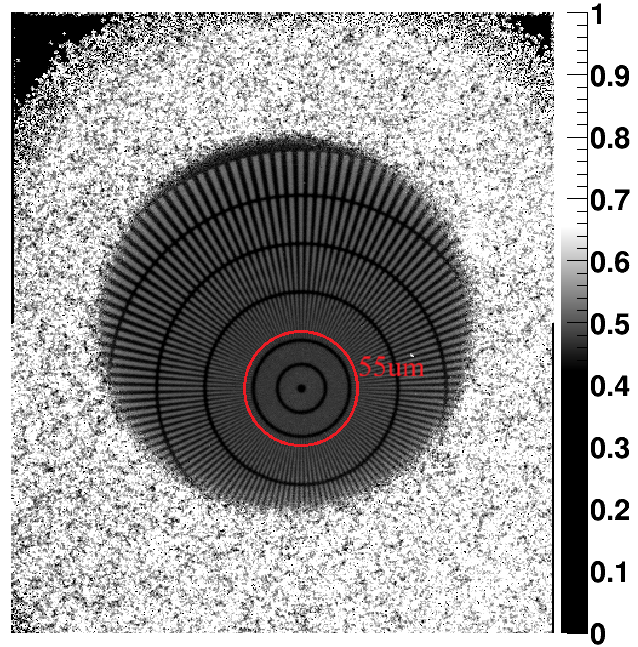} 
    \caption{
    Neutron transmission image of a Siemens Star test sample acquired by the CTPX1 camera under $1\times$ optical magnification. The image is flat-field corrected.
    }
    \label{fig:star_Imaging}
\end{figure}

The imaging capability of the CTPX1 system was qualitatively assessed via neutron transmission imaging using a standard Siemens Star test target.
Figure~\ref{fig:star_Imaging} presents the 2D integrated image acquired by the camera. 
Notably, the image underwent only basic flat-field correction based on open-beam data, without the application of complex clustering or super-resolution algorithms.
As observed in the figure, the spokes of the Siemens Star are clearly resolved, with the limiting spatial resolution primarily determined by the \qty{55}{\micro\meter} physical pixel pitch of the Timepix4 ASIC.
This result confirms that the camera can effectively acquire and reconstruct 2D neutron radiographic images.
For a more detailed quantitative analysis of the spatial resolution of the detector system, we refer the reader to a forthcoming publication by the detector team.\par
\begin{figure}[htbp]
    \centering
    \includegraphics[width=0.6\textwidth]{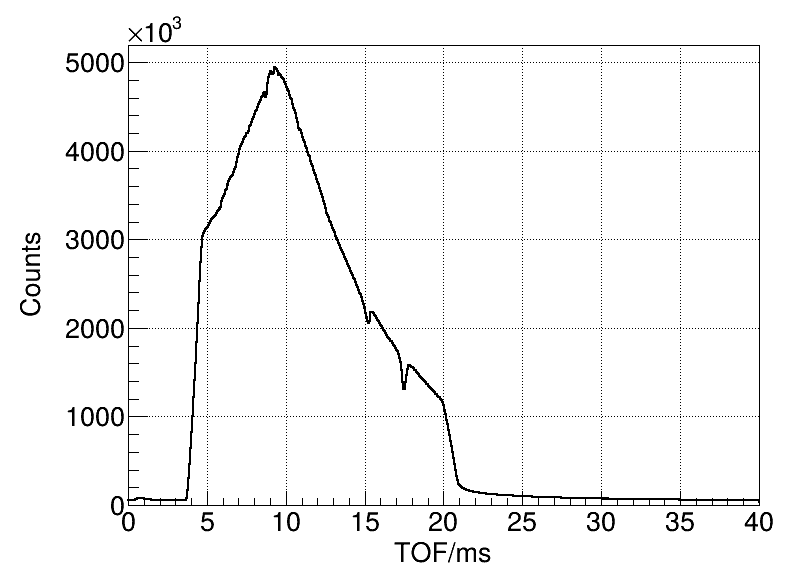} 
    \caption{
    The measured ToF spectrum of the $\gamma$-Fe sample. The plot shows the accumulated neutron counts as a function of flight time (ms). The visible dips in the spectrum correspond to the Bragg diffraction edges, verifying the good temporal resolution.
    }
    \label{fig:ToF_Spectrum}
\end{figure}

To validate the time measurement capability of the system, ToF spectrum were conducted using a polycrystalline $\gamma$-Fe sample. 
Leveraging the high time resolution of the Timepix4 ASIC, the system precisely recorded the ToA time for each detected neutron event relative to the facility's T0 trigger.
Figure~\ref{fig:ToF_Spectrum} displays the measured raw ToF spectrum, plotted as neutron counts versus flight time.
The spectrum exhibits a clear pulse structure with distinct intensity drops corresponding to the Bragg edges of the sample's lattice planes, confirming the camera's time measurement capability.
Together with the previously demonstrated spatial resolution, these results validate the system's suitability for energy-resolved neutron imaging applications.
Similarly, a detailed physical analysis of the Bragg edges, including wavelength conversion, will be presented in the forthcoming publication by the detector team.

\section{Conclusion and outlook}
\label{sec:conclusion}
In this paper, we have presented the design, implementation, and comprehensive validation of CTPX1, a compact, data-driven camera system based on the Timepix4 ASIC, specifically engineered to meet the high readout rate requirements of CSNS-II. 
The system is characterized by its compact form factor, housing high-performance electronics, an low-noise high-voltage bias unit (ripple $< \qty{1}{\milli\volt}$), and a thermal management module.
To fully leverage the bandwidth potential of the Timepix4 ASIC, we developed a firmware architecture based on grouped parallel processing and aggregation, capable of managing the full readout of 16 channel \qty{5.12}{Gbps} serial links. High-flux X-ray testing confirmed that CTPX1 achieves a maximum lossless data throughput of \qty{1.17}{Ghits/s}, corresponding to a bandwidth utilization of approximately \qty{93}{\percent} of the theoretical limit.
The neutron imaging performance of the system was further verified through in-beam experiments at the CSNS BL20 beamline. 
The results confirm that CTPX1 is capable of imaging with a spatial resolution determined by the \qty{55}{\micro\meter} physical pixel pitch of the sensor.
Furthermore, its time-resolved capability was successfully demonstrated by the clear spectral structure observed in the TOF spectrum of a $\gamma$-Fe sample.
In summary, the demonstrated capabilities confirm that CTPX1 is fully competent to meet the experimental requirements for energy-resolved neutron imaging at CSNS-II across varying flux regimes.
Future work will focus on precise system characterization, including energy calibration and time calibration. Additionally, we plan to explore FPGA-based real-time data compression algorithms and scale the architecture to multi-chip tiling arrays to support wider detection areas.

\appendix
\acknowledgments
This work was supported by the National Key R\&D Program of China (Grant No. 2024YFE0110003), the National Natural Science Foundation of China (Grant No. 12227810, Grant No.12305348), Guangdong Basic and Applied Basic Research Foundation (Grant No. 2025A1515011426), Guangdong Provincial Talent Program (Contract No. 2023QN10Z253), and Guangdong Provincial Key Laboratory of Advanced Particle Detection Technology (Grant No. 2024B1212010005).

\bibliography{biblio.bib}

@article{su2016time,
  title={Time-of-flight neutron Bragg-edge transmission imaging of microstructures in bent steel plates},
  url={https://www.sciencedirect.com/science/article/abs/pii/S0921509316309492}, 
  doi={https://doi.org/10.1016/j.msea.2016.08.037}, 
  author={Su, Yuhua and Oikawa, Kenichi and Harjo, Stefanus and Shinohara, Takenao and Kai, Tetsuya and Harada, Masahide and Hiroi, Kosuke and Zhang, Shuoyuan and Parker, Joseph Don and Sato, Hirotaka and others},
  journal={Materials Science and Engineering: A},
  volume={675},
  pages={19--31},
  year={2016},
  publisher={Elsevier}
}

@article{Su_Oikawa_Shinohara_Kai_Horino_Idohara_Misaka_Tomota_2021,   
title={Neutron Bragg-edge transmission imaging for microstructure and residual strain in induction hardened gears},  
url={http://dx.doi.org/10.1038/s41598-021-83555-9},  
doi={10.1038/s41598-021-83555-9},  
journal={Scientific Reports},  
author={Su, Yuhua and Oikawa, Kenichi and Shinohara, Takenao and Kai, Tetsuya and Horino, Takashi and Idohara, Osamu and Misaka, Yoshitaka and Tomota, Yo},  
year={2021},  
month={Feb},  
language={en-US}}

@article{lehmann2014energy,
  title={Energy-selective neutron imaging with high spatial resolution and its impact on the study of crystalline-structured materials},
  author={Lehmann, EH and Peetermans, S and Josic, L and Leber, H and van Swygenhoven, H},
  doi={https://doi.org/10.1016/j.nima.2013.08.065},
  journal={Nuclear Instruments and Methods in Physics Research Section A: Accelerators, Spectrometers, Detectors and Associated Equipment},
  volume={735},
  pages={102--109},
  year={2014},
  publisher={Elsevier}
}

@article{shinohara2020energy,
  doi={https://doi.org/10.1063/1.5136034},
  title={The energy-resolved neutron imaging system, RADEN},
  author={Shinohara, Takenao and Kai, Tetsuya and Oikawa, Kenichi and Nakatani, Takeshi and Segawa, Mariko and Hiroi, Kosuke and Su, Yuhua and Ooi, Motoki and Harada, Masahide and Iikura, Hiroshi and others},
  journal={Review of Scientific Instruments},
  volume={91},
  number={4},
  year={2020},
  publisher={AIP Publishing}
}

@article{CHEN2024169460,
title = {The energy-resolved neutron imaging instrument at the China spallation neutron source},
journal = {Nuclear Instruments and Methods in Physics Research Section A: Accelerators, Spectrometers, Detectors and Associated Equipment},
volume = {1064},
pages = {169460},
year = {2024},
issn = {0168-9002},
doi = {https://doi.org/10.1016/j.nima.2024.169460},
url = {https://www.sciencedirect.com/science/article/pii/S0168900224003863},
author = {Jie Chen and Chaoju Yu and Zhirong Zeng and Haibiao Zheng and Shengxiang Wang and Zhijian Tan and Lufeng Yang and Liyi Wang and Xuekai Zhang}
}

@article{yang2021novel,
  title = {A novel energy resolved neutron imaging detector based on a time stamping optical camera for the CSNS},
journal = {Nuclear Instruments and Methods in Physics Research Section A: Accelerators, Spectrometers, Detectors and Associated Equipment},
volume = {1000},
pages = {165222},
year = {2021},
issn = {0168-9002},
doi = {https://doi.org/10.1016/j.nima.2021.165222},
url = {https://www.sciencedirect.com/science/article/pii/S0168900221002060},
author = {Jianqing Yang and Jianrong Zhou and Xingfen Jiang and Jinhao Tan and Lianjun Zhang and Jianjin Zhou and Xiaojuan Zhou and Wenqin Yang and Yuanguang Xia and Jie Chen and XinLi Sun and Quanhu Zhang and Jiang Li and Zhijia Sun and Yuanbo Chen}
}

@article{yang2024development,
  title = {Development of a dual-mode energy-resolved neutron imaging detector: High spatial resolution and large field of view},
journal = {Nuclear Engineering and Technology},
volume = {56},
number = {7},
pages = {2799-2805},
year = {2024},
issn = {1738-5733},
doi = {https://doi.org/10.1016/j.net.2024.02.042},
url = {https://www.sciencedirect.com/science/article/pii/S1738573324001025},
author = {Wenqin Yang and Jianrong Zhou and Jianqing Yang and Xingfen Jiang and Jinhao Tan and Lin Zhu and Xiaojuan Zhou and Yuanguang Xia and Li Yu and Xiuku Wang and Haiyun Teng and Jiajie Li and Yongxiang Qiu and Peixun Shen and Songlin Wang and Yadong Wei and Yushou Song and Jian Zhuang and Yubin Zhao and Junrong Zhang and Zhijia Sun and Yuanbo Chen},
}

@article{fisher2016timepixcam,
  title={TimepixCam: a fast optical imager with time-stamping},
  author={Fisher-Levine, M and Nomerotski, Andrei},
  doi={10.1088/1748-0221/11/03/C03016},
  journal={Journal of Instrumentation},
  volume={11},
  number={03},
  pages={C03016},
  year={2016},
  publisher={IOP Publishing}
}

@article{nomerotski2019imaging,
  title={Imaging and time stamping of photons with nanosecond resolution in timepix based optical cameras},
  author={Nomerotski, Andrei},
  doi={https://doi.org/10.1016/j.nima.2019.05.034},
  journal={Nuclear Instruments and Methods in Physics Research Section A: Accelerators, Spectrometers, Detectors and Associated Equipment},
  volume={937},
  pages={26--30},
  year={2019},
  publisher={Elsevier}
}

@article{poikela2014timepix3,
  title={Timepix3: a 65K channel hybrid pixel readout chip with simultaneous ToA/ToT and sparse readout},
  author={Poikela, Timepix and Plosila, J and Westerlund, T and Campbell, M and De Gaspari, M and Llopart, X and Gromov, V and Kluit, R and Van Beuzekom, M and Zappon, F and others},
  doi = {10.1088/1748-0221/9/05/C05013},
  journal={Journal of instrumentation},
  volume={9},
  number={05},
  pages={C05013},
  year={2014},
  publisher={IOP Publishing}
}

@inproceedings{fu2019operation,
  title={Operation status and upgrade of CSNS},
  author={Fu, SN and Wang, Sheng and others},
  booktitle={Proceedings, 10th International Particle Accelerator Conference (IPAC2019): Melbourne, Australia},
  year={2019}
}

@article{Fu_2018,
doi = {10.1088/1742-6596/1021/1/012002},
url = {https://dx.doi.org/10.1088/1742-6596/1021/1/012002},
year = {2018},
month = {may},
publisher = {IOP Publishing},
volume = {1021},
number = {1},
pages = {012002},
author = {S N Fu and H S Chen and Y B Chen and L Ma and F W Wang},
title = {CSNS Project Construction},
journal = {Journal of Physics: Conference Series}
}

@article{llopart2022timepix4,
 doi = {10.1088/1748-0221/17/01/C01044},
url = {https://dx.doi.org/10.1088/1748-0221/17/01/C01044},
year = {2022},
month = {jan},
publisher = {IOP Publishing},
volume = {17},
number = {01},
pages = {C01044},
author = {X. Llopart and J. Alozy and R. Ballabriga and M. Campbell and R. Casanova and V. Gromov and E.H.M. Heijne and T. Poikela and E. Santin and V. Sriskaran and L. Tlustos and A. Vitkovskiy},
title = {Timepix4, a large area pixel detector readout chip which can be tiled on 4 sides providing sub-200 ps timestamp binning},
journal = {Journal of Instrumentation}
}

@article{Heijhoff_2022,
doi = {10.1088/1748-0221/17/07/P07006},
url = {https://dx.doi.org/10.1088/1748-0221/17/07/P07006},
year = {2022},
month = {jul},
publisher = {IOP Publishing},
volume = {17},
number = {07},
pages = {P07006},
author = {Heijhoff, K. and Akiba, K. and Ballabriga, R. and van Beuzekom, M. and Campbell, M. and Colijn, A.P. and Fransen, M. and Geertsema, R. and Gromov, V. and Llopart Cudie, X.},
title = {Timing performance of the Timepix4 front-end},
journal = {Journal of Instrumentation}
}

@article{li2025development,
  title={Development of readout electronics for a high-speed event-driven neutron imaging detector based on Timepix4},
  author={Li, Qicai and Liu, Hongbin and Cai, D and Guo, H and Jiang, X and Teng, H and Wang, K and Wang, X and Wang, S and Sun, Z and others},
  doi = {10.1088/1748-0221/20/02/C02031},
  journal={Journal of Instrumentation},
  volume={20},
  number={02},
  pages={C02031},
  year={2025},
  publisher={IOP Publishing}
}

@misc{mpsoc,
author = {{Advanced Micro Devices, Inc.}},
title = {{Zynq UltraScale+ MPSoC Product Selection Guide (XMP104)}},
year = {2022},
url = {},
howpublished = {\url{https://docs.amd.com/v/u/en-US/zynq-ultrascale-plus-product-selection-guidem}},
urldate = {October 21, 2024}
}

@misc{advafab_web,
  author = {{Advafab Oy}},
  title = {{Semiconductor Processing Services}},
  howpublished = {\url{http://www.advafab.com}},
  year = {2024}
}

@misc{SenseFuture_TEC103L,
  author       = {{SenseFuture}},
  title        = {{TEC103L $\pm$3A High-Precision Digital Temperature Controller}},
  howpublished = {\url{https://www.sensefuture.com}},
  year         = {2024}
}

@misc{ADHV4702_datasheet,
  title        = {{ADHV4702-1: 24 V to 220 V, Precision Operational Amplifier Data Sheet}},
  author       = {{Analog Devices Inc.}},
  organization = {Analog Devices, Inc.},
  address      = {Norwood, MA, USA},
  year         = {2019},
  note         = {Rev. A},
  howpublished = {\url{https://www.analog.com/media/en/technical-documentation/data-sheets/adhv4702-1.pdf}}

}

@misc{Keithley_DMM7510,
  title        = {{DMM7510 7.5 Digit Graphical Sampling Multimeter Datasheet}},
  author       = {{Keithley Instruments}},
  organization = {Tektronix, Inc.},
  address      = {Cleveland, OH, USA},
  year         = {2024},
  howpublished = {\url{https://www.tek.com/en/products/keithley/digital-multimeter/dmm7510}},
  url          = {}
}

@misc{hamamatsu,
  author = {{Hamamatsu Photonics K.K.}},
  title = {{150KV MICROFOCUS X-RAY SOURCE}},
  howpublished = {\url{https://www.hamamatsu.com.cn/content/dam/hamamatsu-photonics/sites/documents/99_SALES_LIBRARY/etd/L12161-07_TXPR1023E.pdf}},
  url={},
  urldate = {October 21, 2024},
  year = {2020}
}
\end{document}